\documentstyle[multicol,aps,epsf]{revtex}

\newcommand{\bea}{\begin{eqnarray}}
\newcommand{\beq}{\begin{equation}}
\newcommand{\eea}{\end{eqnarray}}
\newcommand{\eeq}{\end{equation}}

\newcommand{\ch}{{\cal H}}

\newcommand{\lsim}{\vcenter to9pt{\hbox{$<$} \vskip -6pt \hbox{$\sim$}}}

\begin{document}

\title{Do attractive bosons condense?}

\author{N.K. Wilkin, J.M.F. Gunn and R.A. Smith}

\address{School of Physics and Space Research,\\ University of
Birmingham,\\
Edgbaston, Birmingham,\\B15 2TT,\\United Kingdom.}

\date{\today}
\maketitle
\begin{abstract}
Motivated by experiments on bose atoms in traps which have attractive
interactions (e.g. $^7$Li), we consider two models which may be solved
exactly. We construct the ground states subject to the constraint that
the system is rotating with angular momentum proportional to the number
of atoms. In a conventional system this would lead to quantised
vortices; here, for attractive interactions,  we find that the angular
momentum is absorbed by the centre of mass motion. Moreover, the state
is {\it uncondensed} and is an example of a `fragmented' condensate
discussed by Nozi\`eres and Saint James. The same models with {\it
repulsive} 
interactions are fully condensed in the thermodynamic limit. 
\end{abstract}
\pacs{Pacs Numbers: }
\begin{multicols}{2}


One of the most novel aspects of the creation of Bose condensates with
neutral atoms in traps is the possibility of observing a bose gas with
{\it attractive} interactions (negative scattering lengths). The case
of $^7$Li has been studied both experimentally\cite{Brad1,Brad2} and
theoretically. Condensation has been predicted to be stable for a
sufficiently small number of particles or sufficiently weak
interactions\cite{Rup1,kag}. The instability to collapse when these
conditions are not obeyed has also been discussed by several
authors\cite{Baym96,Dodd96,Shur96,Pit96,Bab96}.

In this Letter we show, using two exactly soluble models, that there
may be other possibilities for non-condensed states with attractive
interactions. The states are the `fragmented' condensates discussed by
Nozi\`eres and Saint James\cite{noz0} in the context of excitonic bose
condensates. The possibility of such states emerges from the
realisation\cite{noz} that it is the exchange interaction which causes
bosons with {\it repulsive} interactions to condense into a single
one-particle state, if there are several one-particle ground
states. Conversely for attractive interactions, the exchange term is
negative and may prefer `fragmented'\cite{noz0} condensation into more
than one state if there is a degeneracy (or perhaps if the
interactions are sufficiently strong). Kagan {\em et al.}\cite{kag}
argue that trapped gases with sufficiently large negative scattering
lengths are unstable to the formation of clusters using a somewhat
different argument, but with the same physical origin.

The two models we examine are: particles in a harmonic trap with $L$
quanta of angular momenta with attractive interactions treated as a
degenerate perturbation\cite{TrugKive85}; rotating particles in a
harmonic trap interacting with harmonic
interactions\cite{nucl,Girvin83,Johnson91,bulletinb}. (Both of these
cases have been of interest for {\em
fermions}\cite{TrugKive85,Girvin83}, where rotation is replaced by a
magnetic field and the phenomena are related to the fractional quantum
Hall effect.) Rotation is considered in both cases, partly because the
non-rotating ground state, in the thermodynamic limit, is trivial in
both cases (for different reasons) and partly because the response to
rotation is characteristic of superfluidity in the
system\cite{Leg73,Pines}.


Consider the two-dimensional Hamiltonian
\bea
\label{eq:deg1}
\ch&=&\ch_0+\ch_1\nonumber \\
 \ch_0&=&-\frac{1}{2}\nabla^2 +\frac{1}{2} 
\sum_i {\bf r}_i^2 
 \mbox{ and }\ch_1=\frac{\eta}{2} \sum_{i>j}^N 
\delta({\bf r}_i-{\bf
r}_j),
\eea
 in the limit where the dimensionless coupling is weak, $|\eta|\ll 1$,
so that
the contact interaction can be treated perturbatively. We will now
determine the ground state subject to the constraint that the system
contains $L$ quanta of angular momentum. We should note that the
centre of mass variables will separate in this Hamiltonian because the
trap is harmonic. This will be used below.

The single particle spectrum is usefully expressed \cite{flugge} in
terms of
the angular momentum quantum number, $m$, and the radial quantum number,
$n_{\rm r}$:
\beq
\label{eq:sps}
E = |m| + 2n_{\rm r} + 1
\eeq
in dimensionless units. For the non-interacting case, $\eta = 0$, with
$N$ particles it is clear that to minimise the energy $n_{\rm r}=0$
for all particles.  The angular momentum may be expressed as $L =
\sum_m mN_m$ where $N_m$ is the number of particles in the state
$m$. There is a degeneracy, in general, associated with the choice of
the set $\{N_m\}$.  If the angular momentum is positive, $L>0$, then
because of the modulus signs in Eq.~(\ref{eq:sps}) the energy is
minimised
if one takes only {\em positive} integers for $m$. Then for $0<|\eta|\ll
1$ degenerate perturbation theory should be used with a basis set
consisting of all states $\{N_m\}$, such that $L = \sum_{m =0}^L
mN_m$.

The single particle states with $n_{\rm r} =0$ and $m\ge 0$ are of the
form $z^m \exp(-|z|^2/2)$ where $z=x+iy$ and $m$ is the angular
momentum quantum number.  With the
use of a computer algebra package we have been able to study systems
of up to six particles comprehensively. In addition we can prove that
the form of the ground state holds for an arbitrary number of
particles.
 
We find that in all cases the groundstate for the attractive
interaction ($\eta<0$) is $\psi_{z_c}=z_c^L\exp{\sum_{i=1}^N(
-|z_i|^2)}$ where $z_c=\sum_{i=1}^{N}z_i$, is the centre of mass and
$L$ is the total angular momentum in the system. The contact
interaction energy contribution for the groundstate is independent of
$L$, $\epsilon(N,L)\propto \eta N(N-1)/2$. To prove the form of the
groundstate wavefunction we show that $\psi_{z_c}$ is the unique
eigenfunction of $V=\sum_{i<j}
\delta({\bf r}_i-{\bf r}_j)$ corresponding to its largest eigenvalue,
$\lambda_{\max}$. First we note that $\psi_{z_c}$ is trivially an
eigenfunction of $V$ when $L=0$ since it is the only state in the
$L=0$ subspace. Since the centre of mass coordinate, $z_c$, can be
separated out in the Hamiltonian, it follows that $\psi_{z_c}$ is an
eigenfunction for any $L$.
\par
Let us now work in the basis
\begin{eqnarray}
\label{basis}
\left|m_1,m_2\dots m_N\right>=\prod_{i=1}^N z_i^{m_i}e^{-|z_i|^2/2},
\end{eqnarray}
where $\sum m_i=L$. The matrix elements $\left<m|V|m'\right>$ are 
non-negative, and are positive when $m_i+m_j=m'_k+m'_l$ for some
$i,j,k,l$, and $m_p=m'_q$ for the remaining labels. The coefficients 
$\langle m|\psi_{z_c}\rangle$ are all positive, from which it
follows that the eigenvector $\psi_{z_c}$ belongs to the largest
eigenvalue $\lambda_{max}$. To see this note that $\lambda_{\max}$ can 
be derived from variational principle
\begin{eqnarray}
\label{varia}
\lambda_{\max}=\max{\left\{\frac{\left<\psi|V|\psi\right>}
{\left<\psi|\psi\right>}\right\}}=
\max{\left\{\frac{\sum_{ij}V_{ij}\psi_i\psi_j}
{\sum_i \psi_i^2}\right\}}.
\end{eqnarray}
If we take an eigenvector of $\lambda_{\max}$, and replace all its 
components by their absolute values, the variational functional in
Eq.\ (\ref{varia}) cannot decrease, and so must remain at 
$\lambda_{\max}$. It follows that there is an eigenvector of 
$\lambda_{\max}$ whose components are non-negative; $\psi_{z_c}$ has 
non-zero overlap with this eigenvector, and so must belong to
$\lambda_{\max}$.
\par
To prove non-degeneracy of the eigenspace of $\lambda_{\max}$ we note
that the matrix $V_{ij}$ is ``connected'' in the following way: if we
take a basis vector $\left|i\right>$ and consider all $\left|j\right>$ 
with $V_{ij}>0$, then consider all $\left|k\right>$ with $V_{jk}>0$ and 
so on, this includes all
basis vectors. If the eigenspace of $\lambda_{\max}$ is degenerate then
there must be an eigenvector of $\lambda_{\max}$ whose components are of
both signs (since we can choose this eigenvector to be orthogonal to
$\psi_{z_c}$), and all non-zero (if some components are zero, simply
add on a very small amount of $\psi_{z_c}$). 
The vector made by taking the absolute
value of the latter's components will also be an eigenvector. The
difference in value of the variational functional for the two vectors
can only be zero if the $i$-th and $j$-th components have same sign when
$V_{ij}>0$. By connectedness we see that this means all components must
have the same sign, and hence the eigenspace is non-degenerate. This
completes the proof that $\psi_{z_c}$ is the non-degenerate ground
state of the Hamiltonian Eq.~(\ref{eq:deg1}) with attractive
interaction.

To determine the degree of condensation, if any, the single particle
density matrix, $\rho(z,z^{'*})$ is required for the ground
state. Yang\cite{Yang62} showed that off-diagonal long-range order is
associated with the largest eigenvalue of the density matrix (the
magnitude of the eigenvalue is the fraction condensed) with the
`condensate wavefunction' being the associated eigenvector. The notion
of off-diagonal long-range order is not of such great use for trapped
atoms, but this definition of the {\em condensate} is useful in an
inhomogeneous setting. The single particle density matrix has the form:
\beq
\rho(z,z'^{*})=\frac{1}{Q} \int \prod_{i=2}^N {\rm d} z_i \, {\rm
d}z_i^*\; 
\psi(z,z_2,...z_N) 
\psi^*(z',z_2,..z_N),
\eeq
where $Q$ is the normalisation. On integrating we find,
\beq
\rho(z,z'^{*})= \frac{e^{-|z|^2/2} e^{-|z'|^2/2}}{\pi} 
\sum_{m=0}^L z^m z'^{* m} 
\frac{(N-1)^{L-m} \, L!}{N^L (L-m)!\, m!^2}.
\eeq
Thus the resulting eigenfunctions and eigenvalues for a given $m$ are
\beq 
\psi_m=e^{-\frac{|z|^2}{2}} z^{*m} 
\quad \mbox{and} 
\quad 
\rho_m=\frac{(N-1)^{L-m}\, L!}{N^L\, (L-m)!\, m!}.
\eeq

If we now consider the case of $L=N q$ (which in a conventional
system, {\em e.g.}\ $^4$He would correspond to $q$ vortices) then  if a 
condensate exists its eigenvalue will correspond to  $m=q$. Simplifying
we 
find
\begin{equation}
\rho_q=(1-1/N)^{(q N-q)} \frac{(N q)!}{N^q \, (q N-q)! \; q!},
\end{equation}
which can be rewritten as a Poisson distribution in the limit that
$N\to\infty$. On taking the further limit of $q\to \infty$ the maximal
eigenvalue becomes $\rho_q \sim 1/\sqrt{2
\pi q}$.  However, the eigenvalues of significant weight are
distributed over $q- \sqrt{q} \; \lsim \; m \; \lsim \;
q+\sqrt{q}$. This is clearly not the pronounced peak required for a
condensate and is reminiscent of Nozi\`eres and Saint James'
fragmented condensate\cite{noz0}. Lest this be thought to be misleading
for 
small $q$, we note the following results for $q=1$. We find that the
eigenvalue 
where the putative `condensate' would be, $\rho_1(q=1)=e^{-1}$, that 
$\rho_0(q=1)=e^{-1}$ as well and that $\rho_2(q=1)={{1}\over{2}}e^{-1}$. 
The `condensate' is not singled out as having a uniquely large
eigenvalue. 

The first excited state, $\Phi$, is also of interest, as we find a
rudimentary 
`vortex'. The general form is: 
\beq
\Phi = z_c^{L-2} \sum_{i>j}^N (z_i-z_j)^2 \quad \mbox{with} \quad
 \epsilon_1 \propto N(N-2)/2.
\eeq
Again from symmetry considerations we find that we require a minimum
of 2 quanta of angular momentum in order to produce an excited
state. For L=2, we find that there are two possible states for all $N$
and these correspond to the groundstate and excited state we have
described above. Because of the separation of the centre of mass
variables mentioned above, this is in fact a general result: for 
$\psi_{z_c}=z_c^L$ then an excited state is $\Phi$ (although
we have not proved that it is always the {\em first} excited state).

To determine whether the results from the contact interaction model are
likely 
to be generic or are artifacts (for instance of degenerate perturbation
theory), 
we turn to the second model. The Hamiltonian
\cite{Johnson91,bulletinb} (first discussed in the context of nuclear 
physics\cite{nucl}) describes $N$ bosons with attractive
 harmonic coupling ${\tilde{\Lambda}}>0$)($i$ labels the
particles):
\beq
H=-\frac{\hbar^2}{2 m} \sum_i^N \tilde\nabla_i^2 +\frac{k}{2} \sum_i^N
{\bf 
x}_i^2+
\frac{\tilde{\Lambda}}{4} \sum_{i,j}^N ({\bf x}_i-{\bf x}_j)^2,
 \eeq
where we enforce the symmetry of the wavefunctions at the end of the 
calculation. In
dimensionless units, $\Lambda=\tilde{\Lambda}/k$, $y=(\hbar^2/(m
k))^{1/4} x$ 
and $\nabla=(m k/ \hbar^2)^{1/4} \tilde{\nabla}$ 
\beq
\label{eq:ham2}
\ch=-\frac{1}{2} \sum_i^N\nabla_i^2 +\frac{1}{2} \sum_i^N {\bf y}_i^2+
\frac{\Lambda}{4} 
\sum_{i,j}^N ({\bf y}_i-{\bf y}_j)^2,
\eeq
which upon rearrangement leads to
\beq
\ch=- \frac{1}{2}\sum_i^N\nabla_i^2 +\frac{(1+N \Lambda)}{2} \sum_i^N
{\bf 
y}_i^2-
\frac{\Lambda}{2} \left(\sum_{i}^N {\bf y}_i\right)^2.
\eeq
Here we note that the problem in $d$ dimensions separates into $d$ 
one-dimensional problems. Hence we will now restrict ourselves to 
one-dimension for clarity. 

To determine the degree of condensation we again need to calculate the
single particle density matrix.To do this we change variables to the
centre of mass coordinate, $\zeta_N=1/\sqrt{N} \sum_i^N y_i$, and
$\zeta_i, (i=1,...,N-1)$ which are arbitrary but orthogonal to
$\zeta_N$. The Hamiltonian in these variables is
\beq \ch=-\frac{1}{2} \nabla^2_\zeta +\frac{(1+N \Lambda)}{2} \sum_i^N 
\zeta_i^2-
\frac{N \Lambda}{2}\; \zeta_N^2,
\eeq
This leads to the groundstate wavefunction having the form
\beq
\psi=e^{-\frac{1}{2}(1+N \Lambda)^{1/2} {\bbox{\zeta}}^2
}e^{-\frac{1}{2}
(1-(1+ N \Lambda)^{1/2}) \zeta_N^2},
\eeq
where ${\bbox{\zeta}}=\{\zeta_1,..,\zeta_N\}$. 
The corresponding frequencies are $\epsilon_i=(1+N
\Lambda)^{1/2}/2$ for  $i\ne N$ and $\epsilon_N=1/2$. 
We now consider the model in $d \ge 2$ with attractive interactions in
the fixed angular momentum subspaces. In the groundstate we find that
all angular momentum is absorbed by the centre of mass variable.  This
may be seen by noting that the centre of mass oscillators (associated
with the different components of the centre of mass motion) have a
lower associated frequency than the other oscillators describing
relative motion. The physical interpretation is straightforward:
relative motion requires more energy as work must be performed against
the attractive interactions. (The converse will hold true for
repulsive interactions). Hence, for two-dimensions and $L$ quanta of
angular momentum we can immediately write down the wavefunction,
\beq
\psi=z_c^L e^{-\frac{1}{2}(1+N \Lambda)^{1/2} {\bbox{|z|}^2
}}e^{-\frac{1}{2}
(1-(1+ N \Lambda)^{1/2}) {|z_c|}^2}.
\eeq
 In the thermodynamic limit it can be shown that the contribution to
the single particle density matrix of the exponential term associated
with $z_c$ is negligible. Hence, surprisingly, the density matrix
reduces to that of the contact interaction model. The groundstate
wavefunctions will therefore be the same, as will the properties of
the single particle density matrix.

We shall now show that these systems are condensed (at least under some 
conditions) when the interactions are repulsive. Thus the lack of
condensation 
is not due to peculiarities of the models in general, but of the
attractive 
interactions in particular. 

The repulsive case has also been considered, and will be discussed
elsewhere. We find no vortex lattice, although a Laughlin
like state exists for some special values of $L$.

Firstly consider the contact interaction model when there are $N$ quanta
of 
angular momentum in the 
system. Conventionally there would then be one vortex in the system, and
one 
might expect that the ground state (subject to the constraint of the
amount of 
angular momentum) would be:
\beq
\psi_{L=N}^{\rm mft}= \prod_{i=1}^N\left(z_i e^{-|z_i|^2/2}\right)
\eeq
We conjecture the following form for the CIM:
\beq
\psi_{L=N}^{\rm exact}= 
\prod_{i=1}^N\left([z_1+\cdots+z_n-Nz_i]e^{-|z_i|^2/2}\right)
\label{onev}
\eeq
We have demonstrated that this form is correct by explicit calculation
on 
systems of up to 6 bosons. The physical interpretation of this
wavefunction is 
that the bosons are rotating around the centre of mass, which would be a 
condensate if the centre of mass were a $c$-number. We will now show
that in the 
thermodynamic limit the corrections to full condensation are $O(1/N)$.
Consider 
the density matrix constructed from the wavefunction Eq.~(\ref{onev}):
\newpage
\bea
\rho(z,z'^{*})
=e^{-|z|^2/2}e^{-|z'|^2/2}
\int \left\{\prod_{i=2}^N {\rm d}z_i {\rm d} z_i^*\right\}\nonumber\\
\times
 \left(\omega-\frac{(N-1)}{N}z  \right)
\left(\omega^*-\frac{(N-1)}{N}z'^{*}  \right) \nonumber \\ \times 
\prod_{j=2}^N\left\{\left(\omega+\frac{z}{N}-z_i 
\vphantom{\left(\omega^*+\frac{z'^*}{N}-z^*_j\right)}\right)
\left(\omega^*+\frac{z'^*}{N}-z^*_j\right)e^{-|z_j|^2}\right\}
\eea
where $\omega=1/N \sum_{i=2}^N z_i$.Then to separate the integration
over the 
different $z_i$ we introduce a delta 
function for the centre of mass variable:
\beq
1=\int_{-\infty}^\infty d\omega_xd\omega_y \quad \delta\left(\omega_x- 
\frac{1}{N}\sum_{i=2}^N x_i\right) \delta\left(\omega_y-
\frac{1}{N}\sum_{i=2}^N 
y_i\right) 
\eeq
and use the integral representation
\beq
\delta\left({\bbox{\omega}} - \frac{1}{N}\sum_{i=2}^N {\bf r}_i\right) = 
{{1}\over{(2\pi)^2}} \int d{\bbox{\lambda}}\,e^{i{\bbox{\lambda}}\cdot 
\left({\bbox{\omega}}- \frac{1}{N}\sum_{i=2}^N {\bf r}_i\right)}
\eeq
Upon substitution and integration we find that, in the limit that $N\to
\infty$, 
to accuracy $O(1/N^2)$:
\bea
\rho(z,z'^{*})&=&\sum_n \psi(z) \rho_n \psi^*(z') \nonumber\\
&=&e^{-|z|^2/2}\left(\frac{1}{N}
\,\frac{1}{\sqrt{\pi}}\frac{1}{\sqrt{\pi}}+\left[1-\frac{2}{N}\right]\,\frac{z}
{\sqrt{\pi}}\frac{z'^{*}}{\sqrt{\pi}}\right.\nonumber\\
&+&\frac{1}{N}\,
\left.\frac{z^2}{\sqrt{2 \pi}} \frac{{z'^{*}}^2} {\sqrt{2 \pi}}\right)
e^{-|z'|^2/2}
\eea
Thus there is a condensate with eigenvalue $1-2/N$, in the state $z$,
which is fully condensed in the thermodynamic limit. The corrections
of $1/N$ are in the states $1$ and $z^2$. In addition a Laughlin state
is 
empirically found in small systems to be the ground state for
$L=N(N-1)$.

Turning to the harmonic interactions model, we note that the centre of
mass oscillator has a higher frequency than the others when the
interaction is
repulsive. In that case the other oscillators will be populated in
preference when minimising the energy subject to $L$ (a multiple of
$N$) quanta of angular momentum. Now, the other oscillators are
degenerate and the centre of mass factor is not being multiplied by a
large 
multiple of the centre of mass coordinate, which it was in the
attractive case. 
The latter implies that the centre of mass factor is irrelevant in the
thermodynamic limit. Thus the ground state reduces to a 
single particle form [Eq.~(\ref{eq:sps})] and hence the answer will
be a condensate into the state $z^{L/N}$.

In conclusion we have shown that for a model with degenerate ground
states in the absence of interaction, there is no condensate formed
when weak interactions are incorporated. Consequently, in this
particular case, there is no vortex lattice. In a different model,
which does not have degenerate ground states, we have shown that the
particles are uncondensed when given an extensive quantity of angular
momentum. In both cases the angular momentum of the system resides in
the centre of mass motion, in contrast to the more familiar case of
repulsive interactions. This leads to the general hypothesis:
attractive bosons do not condense in the presence of single particle
degeneracy and their angular momentum resides in the centre of mass
motion. The investigation of {\it rotating} $^7$Li might be fruitful
in exposing an uncondensed `ground state'. Repulsive interactions in
the same models lead to condensed ground states, so showing that
attractive interactions do indeed lead to different physics.

We would like to thank M.W. Long, D.A. Lowe and A.G.B. Triulzi for
useful 
discussions and EPSRC for financial support through grants GR/J35238, 
GR/L28784 and GR/L29156.

\end{multicols}
\end{document}